\documentclass[aps,pra,twocolumn]{revtex4-1}

\usepackage{graphicx}
\usepackage{dcolumn}
\usepackage{bm}
\usepackage{color}
\graphicspath{{figures/}}

\begin{document}
\title{Chaos and $\mathcal{PT}$-symmetry breaking transitions in a driven, nonlinear dimer with balanced gain and loss}

\author{Shiguang Rong}
\altaffiliation{Electronic mail: rong\_shiguang@sina.com}
\affiliation{Department of Physics, Hunan University of Science \& Technology, Xiangtan  411201, China}
\affiliation{Department of Physics, Indiana University Purdue University Indianapolis (IUPUI), Indiana 46202, USA}
\author{Qiongtao Xie}
\altaffiliation{Electronic mail: xieqiongtao@gmail.com}
\affiliation{College of Physics and Electronic Engineering, Hainan Normal University, Haikou 571158, China}
\author{Yogesh N. Joglekar}
\affiliation{Department of physics, Indiana University Purdue University Indianapolis (IUPUI), Indiana 46202, USA}

\begin{abstract}
Dynamics of a simple system, such as a two-state (dimer) model, are dramatically changed in the presence of interactions and external driving, and the resultant unitary dynamics show both regular and chaotic regions. We investigate the non-unitary dynamics of such a dimer in the presence of balanced gain and loss for the two states, i.e. a $\mathcal{PT}$ symmetric dimer. We find that at low and high driving frequencies, the $\mathcal{PT}$-symmetric dimer motion continues to be regular, and the system is in the $\mathcal{PT}$-symmetric state. On that other hand, for intermediate driving frequency, the system shows chaotic motion, and is usually in the $\mathcal{PT}$-symmetry broken state. Our results elucidate the interplay between the $\mathcal{PT}$-symmetry breaking transitions and regular-chaotic transitions in an experimentally accessible toy model.
\end{abstract}

\maketitle

\section{Introduction}
\label{sec:intro}
In recent years,  a special class of the non-hermitian Hamiltonians that are invariant under combined operations of parity and time-reversal ($\mathcal{PT}$) has attracted extensive interest. Starting from the seminal, theoretical works of Bender and co-workers three decades ago~\cite{Bender1}, it has now become clear that open classical systems that are faithfully described by $\mathcal{PT}$-symmetric effective Hamiltonians are of great experimental interest~\cite{Ganainy1}. In quantum mechanics, the requirement of a Hermitian Hamiltonian guarantees the existence of real eigenvalues and a complete set of orthogonal eigenvectors, and thus ensures probability conservation~\cite{Sakurai}. However, Bender and co-workers showed that many non-Hermitian Hamiltonians possess entirely real spectra when the non-Hermiticity is below an energy scale determined by the Hermitian part of the Hamiltonian; the spectrum changes into complex conjugate pairs when the gain-loss strength exceeds this threshold, called the $\mathcal{PT}$-symmetry breaking threshold~\cite{Bender2,Bender3}.This transition from purely real to complex-conjugate spectrum is called $\mathcal{PT}$ -symmetry breaking transition.

After the discovery of $\mathcal{PT}$ symmetric continuum Hamiltonians~\cite{Bender1}, initial efforts were focused on developing a self-consistent quantum theory, i.e. a complex extension of quantum mechanics~\cite{Bender2}, where a new, Hamiltonian-dependent inner product is defined to make the eigenfunctions of the non-Hermitian Hamiltonian orthogonal. These efforts led to significant insights into mathematical properties of pseudo-Hermitian operators with real spectra that are self-adjoint with respect to a non-standard inner product~\cite{mostafa1,mostafa2,mostafa3}. Although the complex extension of quantum mechanics based upon non-Hermitian, $\mathcal{PT}$ symmetric Hamiltonians~\cite{Bender2} is most likely is not a fundamental theory~\cite{Lee1}, classical systems with $\mathcal{PT}$ -symmetric \textit{effective Hamiltonians} have been widely realized. This mapping between Hamiltonians and classical systems is primarily based on the equivalence between the Schr$\ddot{\mathrm{o}}$dinger equation for a non-relativistic particle and the Maxwell equation for the slowly varying envelope of the electric field in the paraxial approximation \cite{Ganainy2,Klaiman}. Resultant experimental examples include optical couplers~\cite{Guo1,Ruter}, microwave billiards~\cite{Bittner}, large-scale temporal lattices~\cite{Regensburger}, microring single-mode lasers~\cite{Feng1,Hodaei1}, and coupled resonators \cite{Peng}. Even in the non-interacting, linear regime, these systems show a wide variety of fascinating behaviors \cite{Lin,Feng2} that are absent in their Hermitian counterparts. When a static, $\mathcal{PT}$-symmetric Hamiltonian is replaced by a time-periodic one, the result is a rich phase diagram of $\mathcal{PT}$ symmetric and $\mathcal{PT}$ broken regions that are determined by the strength of the gain-loss term and the frequency of its temporal modulation~\cite{Joglekar,wang,tonylee}. In particular, in the neighborhood of specific modulation frequencies, the $\mathcal{PT}$ threshold is driven down to zero, thus facilitating the $\mathcal{PT}$ breaking transitions at vanishingly small non-Hermiticity.

The theoretical studies of non-interacting Hamiltonians have been extended to the nonlinear systems~\cite{Suchkov,Konotop} such as dipolar Bose-Einstein condensates~\cite{Li1}, and nonlinear optical~\cite{Abdullaev,Li2} and optomechanical~\cite{Lu} structures. They predict that solitons in the strongly coupled, $\mathcal{PT}$ symmetric systems are stable~\cite{boris1,musslimani}, nonlinear quantum Zeno effects can be observed~\cite{abdullaev2}, and chaos in the $\mathcal{PT}$ symmetric systems~\cite{prosen2010}. Experimental studies have observed optical solitons in $\mathcal{PT}$ symmetric synthetic lattices~\cite{wimmer2015}.

In this paper, we present the dynamics of a $\mathcal{PT}$ symmetric, nonlinear dimer under periodic driving field. At low modulation frequency, the structure of the instantaneous fixed points is analyzed. At intermediate frequency, we find that the dimer exhibits chaotic behavior and $\mathcal{PT}$ symmetry broken phase. The condition for the occurrence of chaos is obtained by means of the Melnikov method. At high modulation frequencies, we find that the driving force strongly renormalizes the dimer coupling constant and therefore strongly modifies its dynamical behavior.

\section{Driven dimer with static gain and loss}
\label{sec:dd}

Consider a $\mathcal{PT}$ -symmetric dimer described by the nonlinear Schr\"{o}dinger equation ($\hbar=1$)
\begin{eqnarray}
\label{eq:one}
i\frac{d\psi_1}{dt} & =& -\frac{\nu}{2}\psi_2-\frac{1}{2}[\varepsilon(t)+i\gamma+\lambda\frac{|\psi_2|^2-|\psi_1|^2}{|\psi_1|^2+|\psi_2|^2}]\psi_1,\\
\label{eq:two}
i\frac{d\psi_2}{dt}& =& -\frac{\nu}{2}\psi_1+\frac{1}{2}[\varepsilon(t)+i\gamma+\lambda\frac{|\psi_2|^2-|\psi_1|^2}{|\psi_1|^2+|\psi_2|^2}]\psi_2.
\end{eqnarray}
Here $\nu$ is the coupling between the two sites of the dimer that, in the absence of all other terms,
leads to Rabi oscillations. We use it to set the frequency and time scale in the rest of the paper. $\gamma>0$ is the strength of the balanced gain and loss potential, which, in the absence of all other terms, leads to exponential decay on the first site and exponential amplification of the second site. $\varepsilon(t)$ is an external, Floquet drive characterized by an amplitude $A$ and frequency $\omega$, i.e. $\varepsilon(t) = A\sin (\omega t) =-\varepsilon(-t)$ , and $\lambda$ is the antisymmetric nonlinearity that is proportional to the dimer polarization. When $\lambda = 0$, this system shows a sequence of $\mathcal{PT}$ -symmetric transitions when the driving frequency $\omega$ is changed at  a fixed value of driving strength $A$~\cite{Luo,Li3,Roberto}. It is convenient to cast Eqs.(\ref{eq:one})-(\ref{eq:two}) in a matrix form by defining an effective, state-dependent Hamiltonian as $i\partial_t|\psi(t)\rangle= H_{\mathrm{eff}}(t)|\psi(t)\rangle$, where the $2\times2$ Hamiltonian is
\begin{equation}
\label{eq:heff}
H_{\mathrm{eff}}(t)= -\frac{\nu}{2}\sigma_x-\frac{1}{2}\left[\varepsilon(t) + i\gamma - \lambda Z(t) \right]\sigma_z,
\end{equation}
where $\sigma_x,\sigma_z$ are the standard Pauli matrices, and $Z(t)\equiv\langle\psi(t)|\sigma_z|\psi(t)\rangle/\langle\psi(t)|\psi(t)\rangle$ is the time-dependent polarization of the dimer. Note that, by definition, the dimer polarization is a bounded, real function, $-1\leq Z(t)\leq 1$. It is straightforward to check that $H_{\mathrm{eff}}$ commutes with the $\mathcal{PT}$ operator where $\mathcal{P} = \sigma_x$ exchanges the first site of the dimer with the second, and $\mathcal{T}$: $t\rightarrow-t, i\rightarrow-i$ is the time-reversal operation. Therefore, the driven, nonlinear, gain-loss dimer model is $\mathcal{PT}$ symmetric.

Due to the presence of the gain and loss terms $\pm i\gamma$, the evolution generated by the time-dependent effective Hamiltonian $H_{\mathrm{eff}}$ is not unitary. The norm of the dimer wavefunction $n(t)\equiv\langle\psi(t)|\psi(t)\rangle= |\psi_1(t)|^2 + |\psi_2(t)|^2$ is not conserved and depends on the dimer polarization, i.e. $dn/dt=-\gamma n(t)Z(t)\neq0$. Therefore, starting from a normalized initial state $|\psi(0)\rangle$) on the Bloch sphere, the time-evolved state does not remain confined to it. We separate this motion of $|\psi(t)\rangle$ into the dynamics of its norm $n(t)$ and its projection onto the Bloch sphere at every instance of time, and consider a new, scaled state
\begin{equation}
\label{eq:psiscaled}
|\psi'(t)\rangle\equiv\frac{|\psi(t)\rangle}{\sqrt{\langle\psi(t)|\psi(t)\rangle}}=\frac{|\psi(t)\rangle}{\sqrt{n(t)}},
\end{equation}
that is normalized at all times and satisfies a differential equation $i\partial_t|\psi'(t)\rangle= H'_{\mathrm{eff}}|\psi'(t)\rangle$ with a scaled effective Hamiltonian $H'_{\mathrm{eff}}$ given by
\begin{equation}
\label{eq:heffscaled}
H'_{\mathrm{eff}} =+\frac{i\gamma}{2}Z(t)-\frac{\nu}{2}\sigma_x - \frac{1}{2}[\varepsilon(t)+i\gamma - \lambda Z(t)]\sigma_z,
\end{equation}
We would like to emphasize that although $H'_{\mathrm{eff}}$ is not Hermitian, it conserves the norm of the state, i.e. $\partial_t\langle\psi'(t)|\psi'(t)\rangle = 0$. The equation of motion for the scaled state $|\psi'(t)\rangle = (\psi_1',\psi_2')^T$ is simplified by expressing it in terms of the polarization $Z(t)$ and two phases $0\leq\theta_1(t),\theta_2(t)\leq 2\pi$,
\begin{equation}
\label{eq:psiscaled12}
\psi'_{1,2}(t)\equiv\sqrt{\frac{1\pm Z(t)}{2}}e^{i\theta_{1,2}(t)}.
\end{equation}
The Schr\"{o}dinger equation $i\partial_t|\psi'(t)\rangle=H'_{\mathrm{eff}}(t)|\psi'(t)\rangle$ for the driven, nonlinear, $\mathcal{PT}$ -symmetric dimer then becomes
\begin{eqnarray}
\label{eq:z}
\partial_t Z&  =& -\nu\sqrt{1-Z^2}\sin\theta - \gamma(1-Z^2),\\
\label{eq:theta}
\partial_t\theta & =& -\varepsilon(t) + \lambda Z + \nu\frac{Z}{\sqrt{1-Z^2}}\cos\theta,
\end{eqnarray}
where $\theta(t)\equiv\theta_2(t)-\theta_1(t)$ is the phase difference between the wavefunction weights on the two sites. In the following section, we investigate the properties of these two equations for a sinusoidal drive $\varepsilon(t)=A\sin(\omega t)$ as a function of the amplitude $A$ and the driving frequency $\omega$ across the entire frequency range.

\begin{figure*}
\begin{center}
\includegraphics[width=\columnwidth]{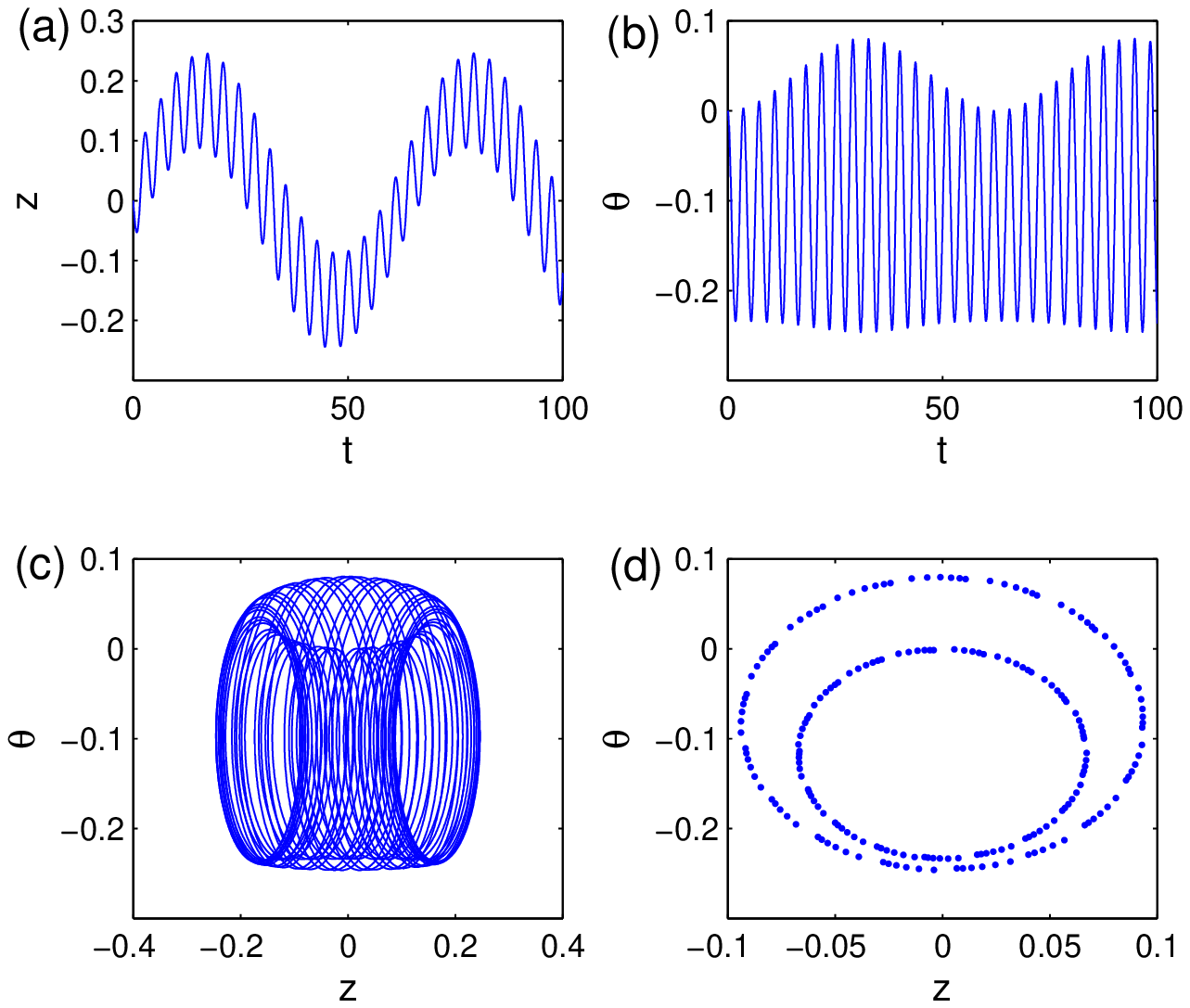}
\includegraphics[width=\columnwidth]{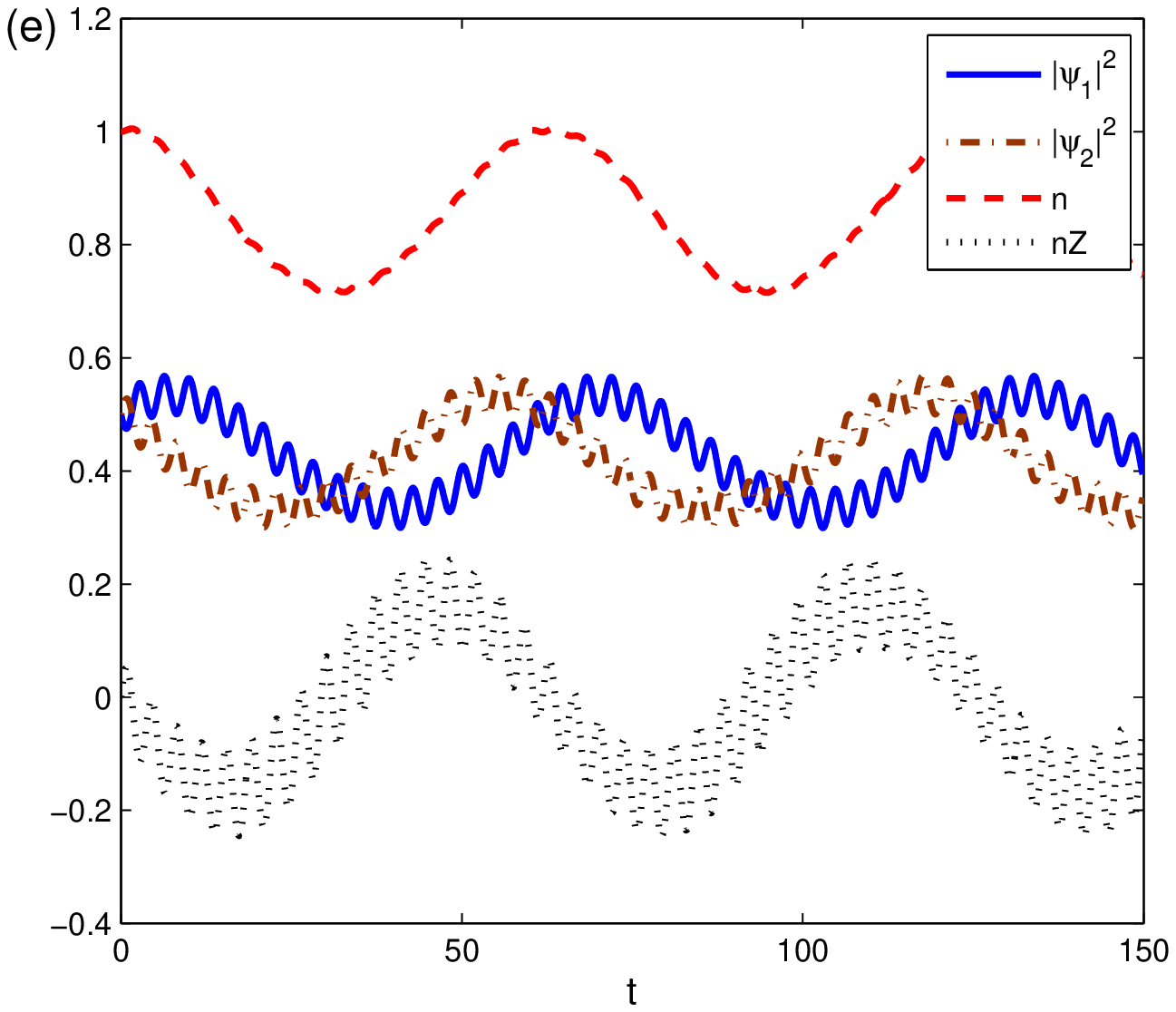}
\caption{Regular behavior of a $\mathcal{PT}$ symmetric dimer with $\gamma/\nu=0.1$ and strong nonlinearity $\lambda/\nu=2$, driven by an external force with moderate amplitude $A/\nu=0.5$ and low frequency $\omega/\nu=0.1$. Temporal evolution of the dimer polarization $Z(t)$ (a) and phase difference $\theta(t)$ (b) shows oscillatory behavior with two distinct frequency components $\nu$ and $\omega=\nu/10$. The phase-space portrait (c) and the Poincare section (d) in the $Z-\theta$ plane show that the dimer evolution is regular. (e) The temporal evolution of norm of the state $n(t)$, the wavefunction weights on the two sites $|\psi_{1,2}|^2$, and their difference shows periodic behavior that is characteristic of a $\mathcal{PT}$ symmetric phase.}
\label{fig:one}
\end{center}
\end{figure*}

\section{Low-frequency driving: $\omega/\nu \ll 1$}
\label{sec:lowf}
In the static-driving limit, we consider the instantaneous fixed points of Eqs.(\ref{eq:z})-(\ref{eq:theta}). The steady-state dimer polarization $Z_f$ satisfies the equation
\begin{eqnarray}
(\gamma^2+\lambda^2)Z_f^4-2\varepsilon\lambda Z_f^3+(\nu^2+\varepsilon^2-\gamma^2-\lambda^2)Z_f^2\nonumber\\
\label{eq:zf}
+2\varepsilon\lambda Z_f-\varepsilon^2=0.
\end{eqnarray}
In the absence of an external drive, $\varepsilon=0$, Eq.(\ref{eq:zf}) becomes a biquadratic and its fixed points are analytically obtained,
\begin{equation}
\label{eq:zfvalues}
Z_f=\left\{0,0,\pm\sqrt{1-\frac{\nu^2}{\gamma^2+\lambda^2}}\right\}.
\end{equation}
The corresponding, doubly-degenerate, steady-state phase difference values $\theta_f$ are given by
\begin{equation}
\label{eq:thetafvalue}
\cos\theta_f=\left\{-\frac{\lambda}{\nu},-\frac{\lambda}{\sqrt{\lambda^2+\nu^2}}\right\}.
\end{equation}
This static-limit analysis can be extended to the case $\varepsilon\neq0$ in a straight forward manner. At low frequencies, we numerically obtain the temporal evolution of the dimer by solving Eqs.(\ref{eq:z})-(\ref{eq:theta}) with given initial conditions. Figure~\ref{fig:one} shows typical results for a dimer with
gain-loss strength $\gamma/\nu=0.1$ and a strong nonlinearity $\lambda/\nu=2$, driven by a moderate strength, low-frequency external force with $A/\nu=0.5$ and $\omega/\nu=0.1$. The dimer is initially in a symmetric state, $|\psi(0)\rangle=(1,1)^T/\sqrt{2}$ or, equivalently, $Z(0)=0=\theta(0)$. Figure~\ref{fig:one}a,b show that the dimer polarization $Z(t)$ and phase difference $\theta(t)$ oscillate periodically with two dominant frequency components, namely $\nu$ and $\omega$. Figure~\ref{fig:one}c,d are the phase-space portrait and Poincare sections of the corresponding time evolution. They show that the dimer has a regular motion when driven at low frequencies. In Fig.~\ref{fig:one}e, we plot the norm $n(t)$ of the state vector $|\psi(t)\rangle$, the weights $|\psi_{1,2}(t)|^2$ on the two sites, and their difference $\langle\psi(t)|\sigma_z|\psi(t)\rangle=n(t)Z(t)$. The norm $n(t)$ shows oscillatory behavior that is a hallmark of the $\mathcal{PT}$ symmetric phase, with two frequency scales, as do the other quantities. These results show that in the low driving frequency regime, the driven, nonlinear, $\mathcal{PT}$ symmetric dimer is in the $\mathcal{PT}$ symmetric phase and has regular, non-chaotic dynamics.

Due to strong nonlinearity in this system, an exhaustive or analytical investigation of the $\mathcal{PT}$ symmetric phase diagram as a function of the four dimensionless parameters, i.e. the amplitude of the drive $A/\nu$, the frequency of the drive $\omega/\nu$, the strength of nonlinearity $\lambda/\nu$, and the gain-loss strength $\gamma/\nu$ is virtually impossible. Therefore, in this work, we focus primarily on the regime with strong nonlinearity and moderate external drive. We note that in the linear case ($\lambda=0$), this characterization can be analytically carried out~\cite{Luo,Li3}, and in the static case, a variety of integrable $\mathcal{PT}$ symmetric dimer models have been studied in the literature~\cite{pickton,Bara0,Bara1,khare}.

\begin{figure*}
\centering
\includegraphics[width=\columnwidth]{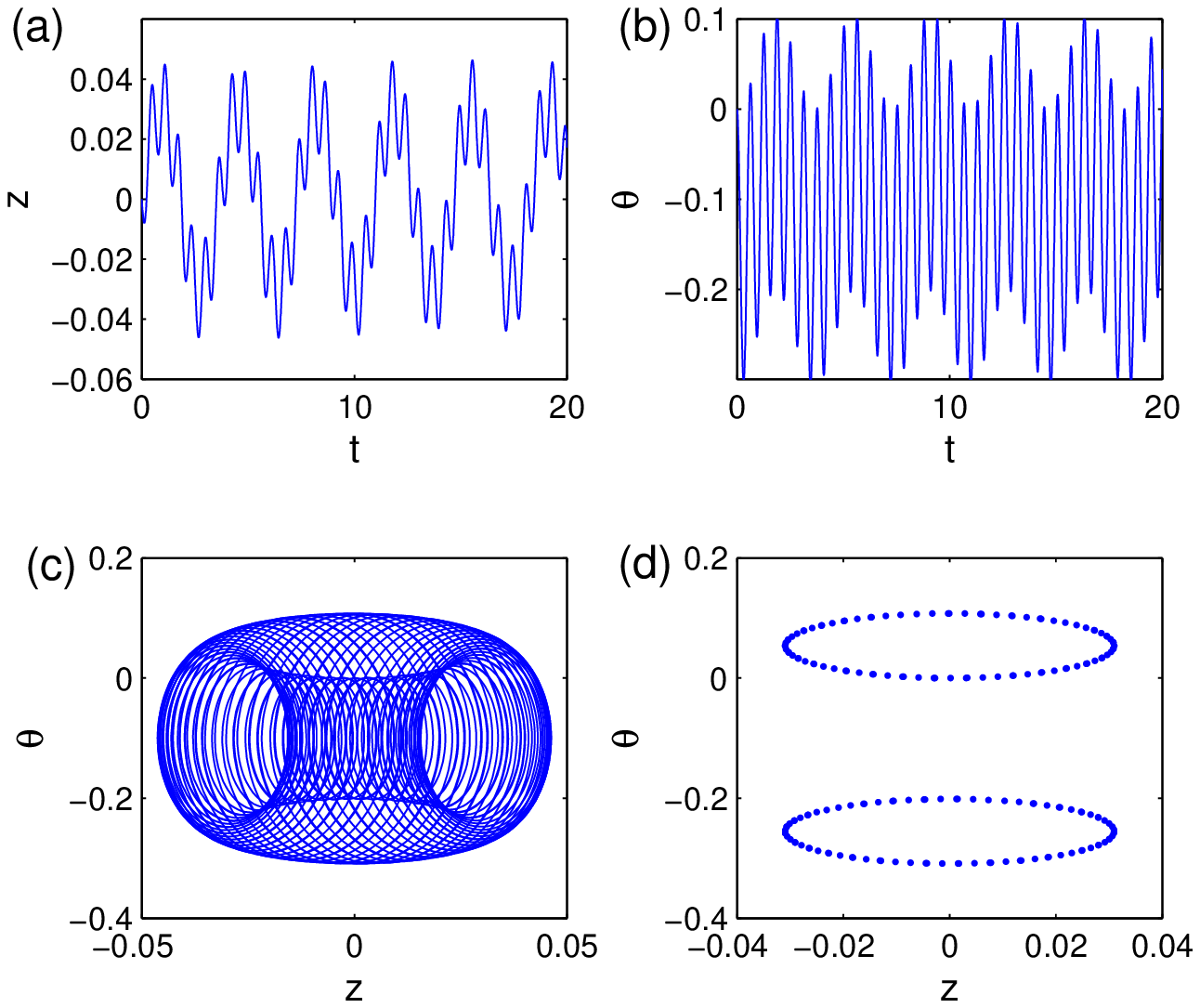}
\includegraphics[width=\columnwidth]{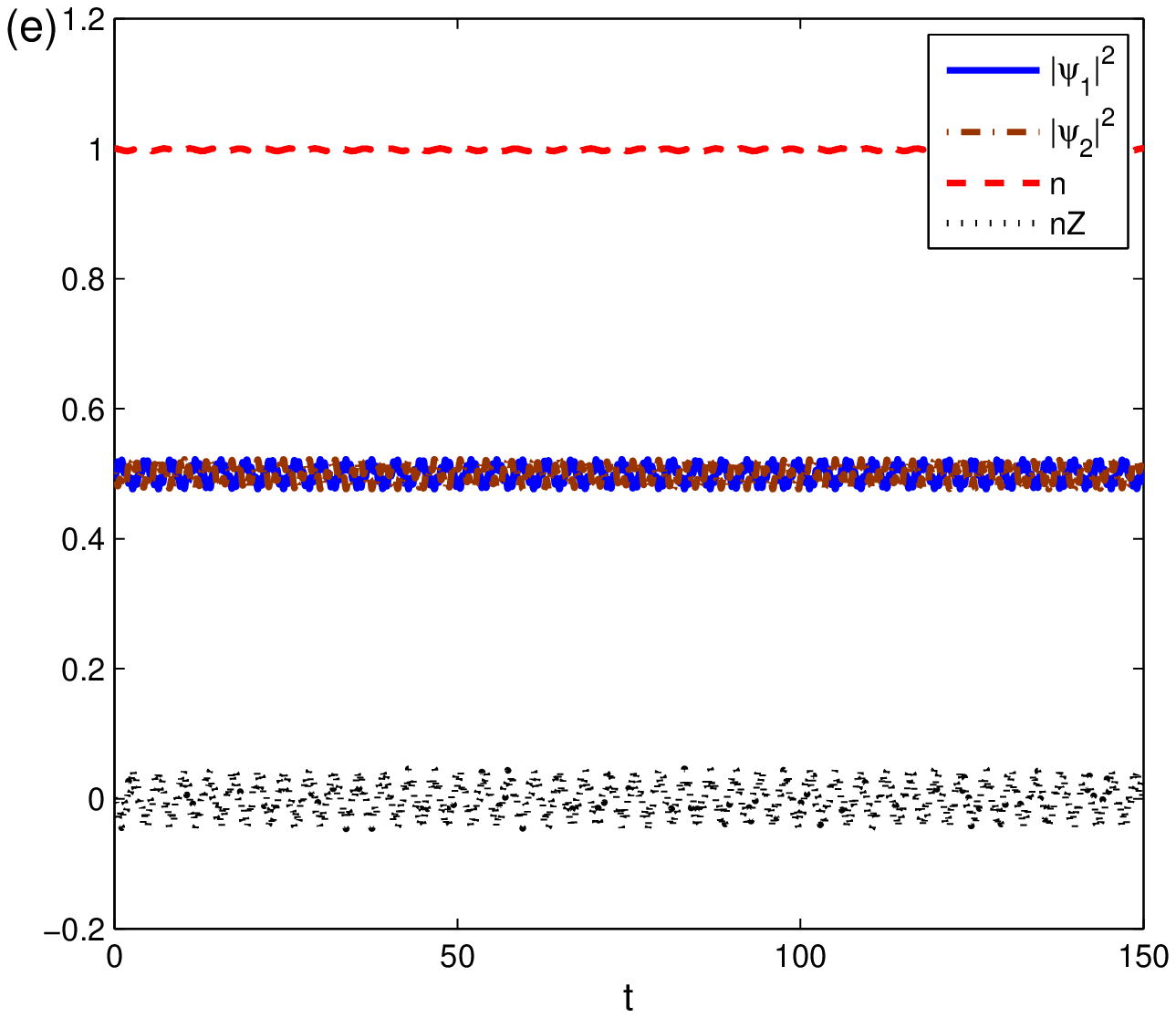}
\caption{Quasi-static, regular motion of the $\mathcal{PT}$ symmetric dimer with $\gamma/\nu=0.1$ and strong nonlinearity $\lambda/\nu=2$, driven by an external force with moderate amplitude $A/\nu=0.5$ and high frequency $\omega/\nu=10$. The effective coupling amplitude is essentially equal to the bare coupling, i.e. $\nu_{\mathrm{eff}}/\nu=J_0(A/\omega)=1$. The temporal evolution of the dimer polarization $Z(t)$ (a) and phase difference $\theta(t)$ (b) shows oscillatory behavior with two frequencies $\nu$ and $\omega=10\nu$. The phase space portrait (c) and the Poincare section (d) in the $Z-\theta$ plane show that the dimer dynamics is regular. (e) The temporal evolution of norm of the state $n(t)$, the wavefunction weights on the two sites $|\psi_{1,2}|^2$, and their difference shows periodic behavior that is characteristic of a $\mathcal{PT}$ symmetric phase.}
\label{fig:two}
\end{figure*}

\section{High-frequency regime: $\omega/\nu \gg 1$}
\label{sec:highf}
When the external drive frequency is much larger than the Rabi frequency of the dimer, we can separate the dynamics of $|\psi'(t)\rangle$ into the high-frequency contribution and a slowly varying field, i.e.
\begin{equation}
\label{eq:varphi}
|\psi'(t)\rangle=\exp\left[\pm\frac{i}{2}\sigma_z\int_{\pi/2}^t dt'\varepsilon(t')\right]|\varphi(t)\rangle.
\end{equation}
Note that the dimer polarization is solely determined by the slowly varying field, $Z(t)=\langle\psi'|\sigma_z|\psi'\rangle=\langle\varphi|\sigma_z|\varphi\rangle$. The equation of motion for the slowly varying field is given by
\begin{eqnarray}
i\partial_t|\varphi(t)\rangle & = & \frac{i\gamma}{2}Z|\varphi(t)\rangle-\frac{1}{2}\left[i\gamma-\lambda Z\right]\sigma_z|\varphi(t)\rangle\nonumber\\
& & -\frac{1}{2}\left[ \nu(t) \sigma_{+} + \nu^*(t)\sigma_{-}\right]|\varphi(t)\rangle,
\end{eqnarray}
where $\nu(t)=\nu\exp(iA\cos\omega t/\omega)$ is the complex coupling amplitude and $\sigma_{\pm}=(\sigma_x\pm i\sigma_y)/2$. In the high-frequency limit, ignoring the higher-order Bessel functions in the expansion of the exponential-cosine gives the following effective, time-independent Hamiltonian for the $\varphi$ field,
\begin{equation}
\label{eq:Hs}
H'_s=+\frac{i\gamma}{2}Z-\frac{\nu J_0(A/\omega)}{2}\sigma_x-\frac{1}{2}\left[i\gamma-\lambda Z\right]\sigma_z.
\end{equation}
Comparison of Eq.(\ref{eq:Hs}) with Eq.(\ref{eq:heffscaled}) shows that, to first approximation, the high-frequency Hamiltonian behaves like a nonlinear dimer with no drive ($\varepsilon=0$) and a smaller, effective coupling $\nu\rightarrow \nu_{\mathrm{eff}}=\nu J_0(A/\omega)$. Since $\nu_{\mathrm{eff}}$ can be made arbitrarily small or driven to zero by appropriate choice of $A/\omega$, the system can be driven from $\mathcal{PT}$ symmetric phase to $\mathcal{PT}$ broken phase and back~\cite{Luo}.

In Fig.~\ref{fig:two} we show the typical results for the temporal evolution of such a dimer with the same initial state as in Fig.~\ref{fig:one}, but a high driving frequency $\omega/\nu=10$. We note that for these parameters, the effective coupling is essentially equal to the bare coupling, i.e. $\nu_{\mathrm{eff}}=\nu J_0(A/\omega)=\nu$. Figure~\ref{fig:two}a, b show that the dimer polarization and phase differences both evolve periodically with two distinct frequencies, and Fig.~\ref{fig:two}c, d show that the dynamics are regular when the dimer is driven by a high-frequency field. The temporal evolution of the norm of the state $n(t)$, on-site weights, and their differences, Fig.~\ref{fig:two}e, show oscillatory behavior consistent with the $\mathcal{PT}$ symmetric phase. In particular, the smallness of oscillations in Fig.~\ref{fig:two}e shows that the system is deep in the $\mathcal{PT}$ symmetric phase, or almost Hermitian.

\section{ Intermediate-frequency driving}
\label{sec:inter}

\begin{figure*}
\centering
\includegraphics[width=\columnwidth]{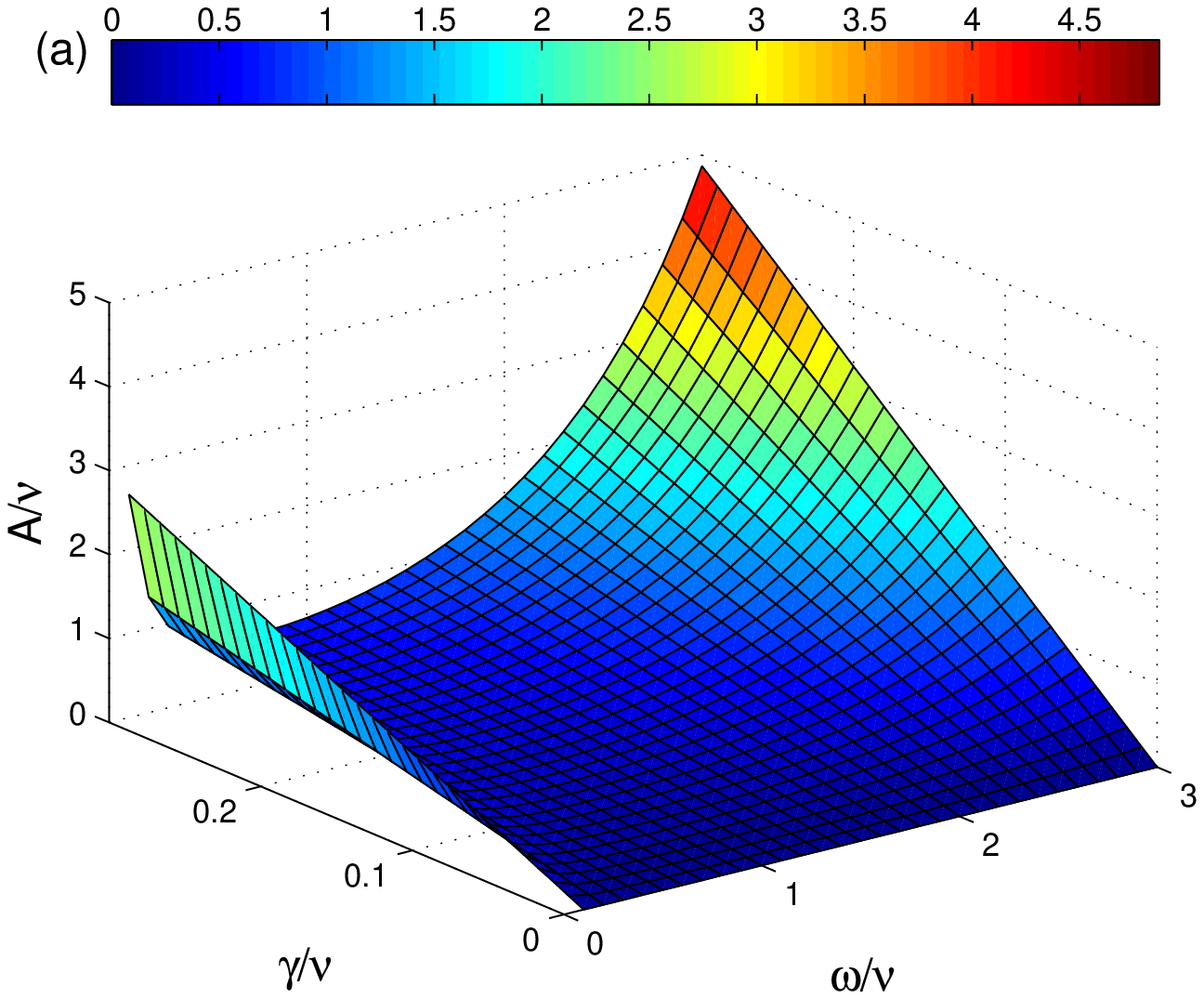}
\includegraphics[width=\columnwidth]{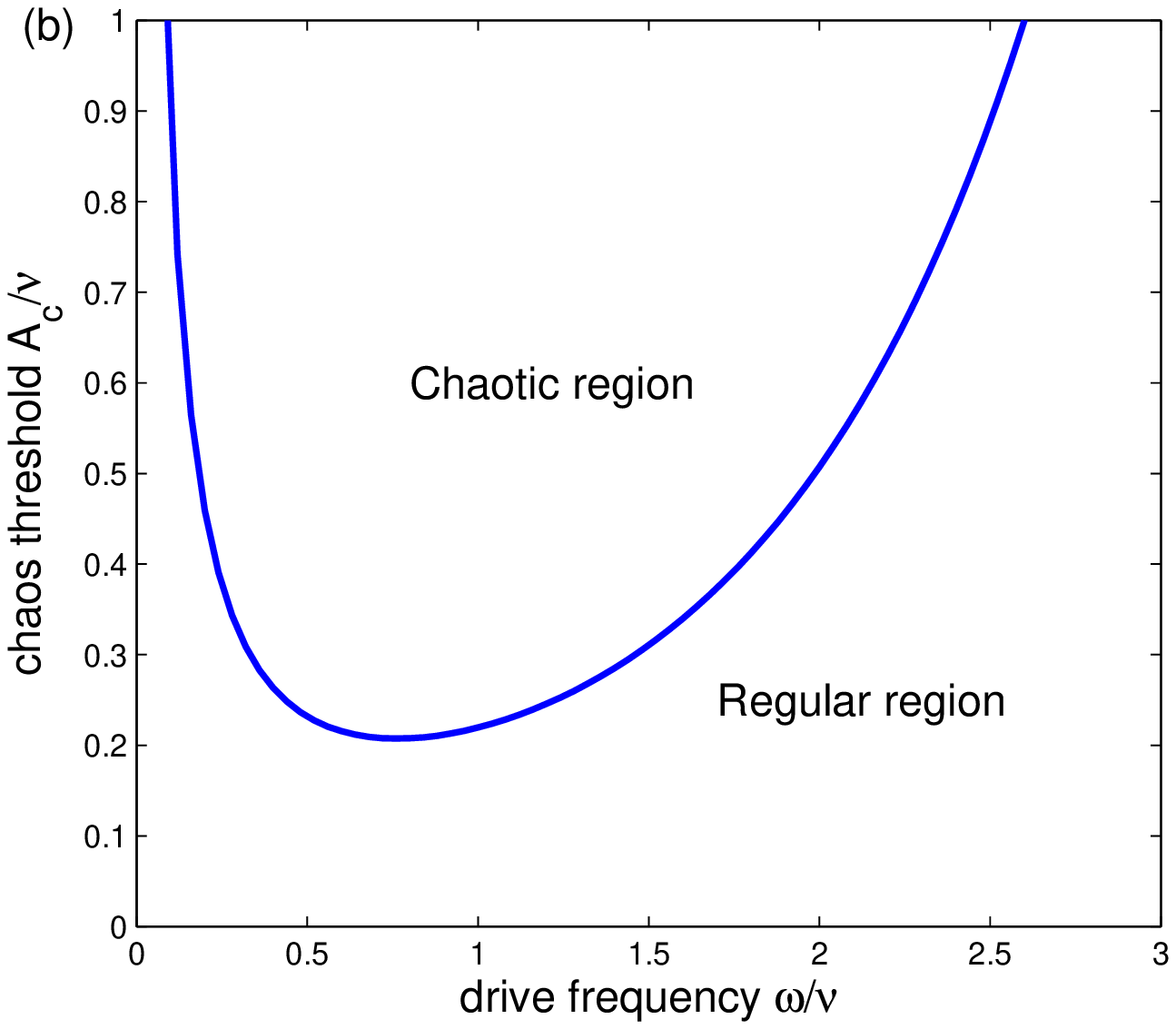}
\caption{Chaos threshold amplitude $A_c$ as a function of the gain-loss strength $\gamma$ and drive frequency $\omega$, for a $\mathcal{PT}$ symmetric dimer with strong nonlinearity, $\lambda/\nu=2$. (a) $A_c(\gamma,\omega)$ shows a linear-in-$\gamma$ dependence, Eq.(\ref{eq:ac}), and a marked minimum at intermediate frequencies $\omega/\nu\sim1$. (b) A cut at $\gamma/\nu=0.1$ shows rapidly divergent behavior at low and high drive frequencies.}
\label{fig:three}
\end{figure*}

At last, we consider the most interesting case, namely that of intermediate modulation frequencies $\omega\sim\nu$. When there is no external drive or gain/loss potential, the resulting system of equations
\begin{eqnarray}
\label{eq:zexact}
\partial_t Z(t) & = & -\nu \sqrt{1-Z^2}\sin\theta\equiv f_1,\\
\label{eq:thetaexact}
\partial_t\theta(t) & = & \lambda Z+\nu\frac{Z}{\sqrt{1-Z^2}}\cos\theta\equiv f_2,
\end{eqnarray}
is integrable~\cite{Raghavan}. The equations (\ref{eq:zexact})-(\ref{eq:thetaexact}) can be written in the Hamiltonian form~\cite{Abdullaev3} with Hamiltonian
\begin{eqnarray}
\label{eq:Hamiltonian}
H=\frac{\lambda}{2}Z^2-\nu\sqrt{1-Z^2}\cos\theta.
\end{eqnarray}
By introducing an effective potential $V(Z)=Z^2(\lambda^2/2-\lambda H/2)+ \lambda^2Z^2/8$, we find that the field $Z(t)$ satisfies Newton's second law, i.e. $\partial_{tt}Z=-\partial V/\partial Z$, and its total energy is given by 
\begin{eqnarray}
\label{eq:totalenegy}
E_0=\frac{1}{2}\left(\frac{\partial Z}{\partial t}\right)^2+V(Z)=\frac{1}{2}\left(\nu^2-H^2\right).
\end{eqnarray}
The separatrix solution of Eqs.(\ref{eq:zexact})-(\ref{eq:thetaexact}), i.e. the $E_0=0$ case, is given by
\begin{eqnarray}
\label{eq:z2exact}
Z_s(t)& =& \frac{2a}{\lambda}\mathrm{sech}(at),\\
\label{eq:theta2exact}
\sin^2[\theta_s(t)] & = & \frac{\lambda^2a^2\mathrm{sech}^2(at)\mathrm{tanh}^2(at)}{\lambda^2-4a^2\mathrm{sech}^2(at)},
\end{eqnarray}
where $a=\sqrt{\nu\lambda-\nu^2}$. When this exact solution is subject to a periodic perturbation, its stability analysis is carried out via Melnikov method~\cite{gucken}. We start with the equation for the two-component vector ${\bf z}\equiv (Z,\theta)^T$,
\begin{equation}
\label{eq:bfz}
\partial_t{\bf z}(t)={\bf f}(t)+\epsilon{\bf g}(t),
\end{equation}
where ${\bf f}\equiv (f_1,f_2)^T$, $\epsilon\ll 1$ is a dimensionless small parameter, and ${\bf g}(t)\equiv (g_1,g_2)^T$ with
\begin{eqnarray}
\label{eq:g1}
g_1(t)& = & -\gamma\sqrt{1-Z^2}/\epsilon,\\
\label{eq:g2}
g_2(t)& = &-A\sin\omega t/\epsilon.
\end{eqnarray}
The Melnikov function $M(t_0)$ can be analytically calculated from Eqs.(\ref{eq:z2exact})-(\ref{eq:theta2exact}), and gives
\begin{eqnarray}
M(t_0 ) & = & -\int_{ - \infty }^\infty {\bf f}(t,t_0)\wedge{\bf g}(t,t_0) dt,\nonumber\\
\label{eq:melnikov}
& = & -(\gamma F_1 + F_2 A\cos\omega t_0)/\epsilon.
\end{eqnarray}
where ${\bf f}\wedge{\bf g}\equiv \left(f_1g_2-f_2g_1\right)$, the two auxiliary functions are $F_1=\pi(\lambda-\nu)(4\lambda-\nu)/2\lambda^2$ and $F_2=(2\pi\omega/\lambda)\mathrm{sech}\left(\pi\omega/2\sqrt{\lambda\nu-\nu^2}\right)$, and $t_0$ is the initial time. The condition for the onset of classical chaos is given by $M(t_0)=0$, or, equivalently, the drive amplitude must exceed the chaos threshold, i.e. $A\geq A_c$ where
\begin{equation}
\label{eq:ac}
A_c=\gamma\frac{F_1(\lambda,\nu)}{F_2(\omega,\lambda,\nu)}.
\end{equation}

Figure~\ref{fig:three}a shows the dimensionless chaos threshold amplitude $A_c/\nu$ as a function of
the gain-loss strength $\gamma$ and the drive frequency $\omega$, at a strong nonlinearity $\lambda/\nu=2$. In the absence of gain and loss, $\gamma=0$, the chaos threshold is zero, meaning the nonlinear, driven dimer is always in the chaotic region. As the gain-loss strength is increased, the chaos threshold is large at both high and low frequencies, but is suppressed at intermediate driving frequencies. Figure~\ref{fig:three}b is a vertical cut at $\gamma/\nu=0.1$, showing a rapidly divergent chaos threshold in the static limit and in the high-frequency regime. We note that the linear-in-$\gamma$ dependence of the chaos threshold, seen in Fig.~\ref{fig:three}a, is a reflection of Eq.(\ref{eq:ac}).

\begin{figure}[h]
\centering
\includegraphics[width=0.49\columnwidth]{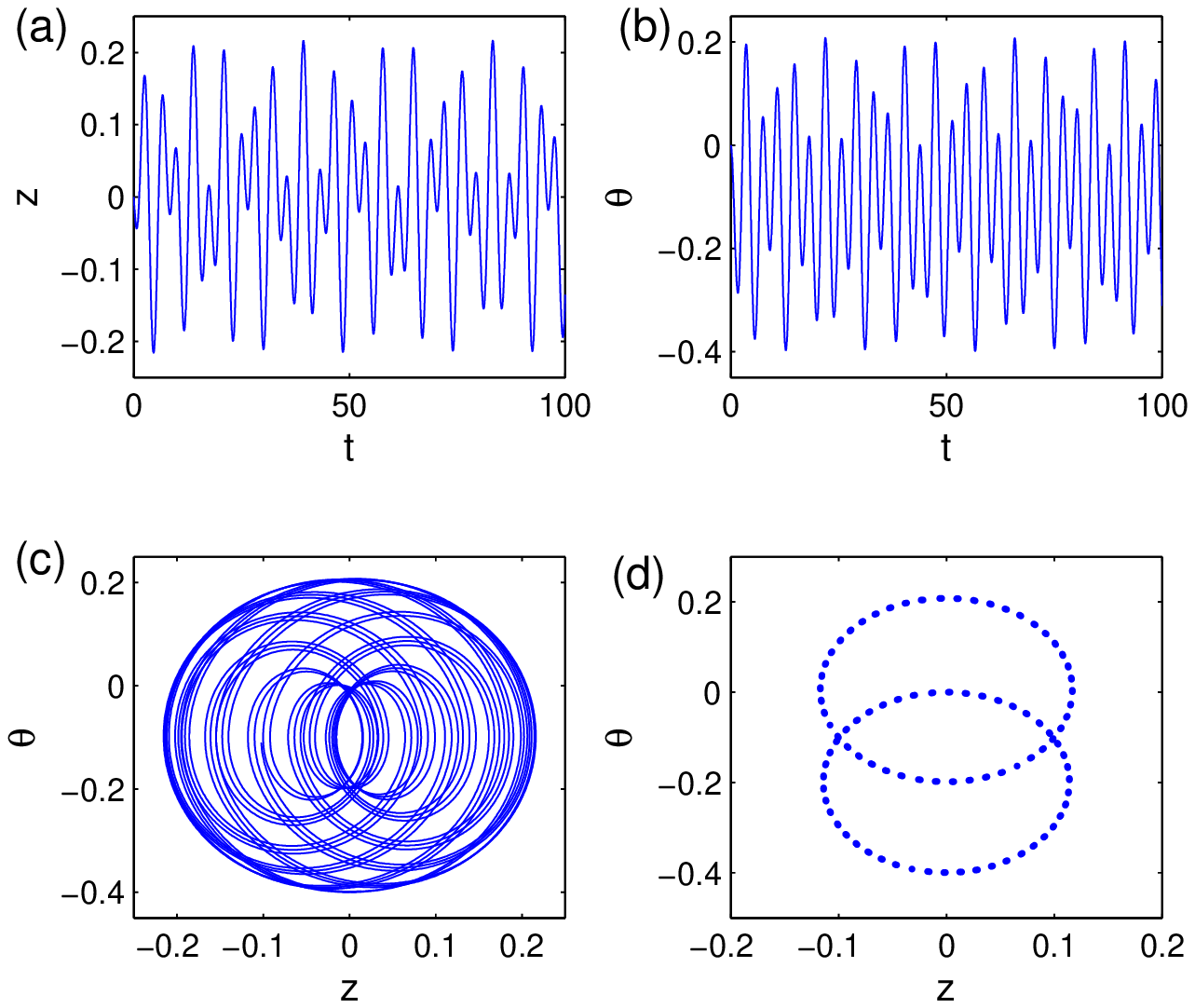}
\includegraphics[width=0.49\columnwidth]{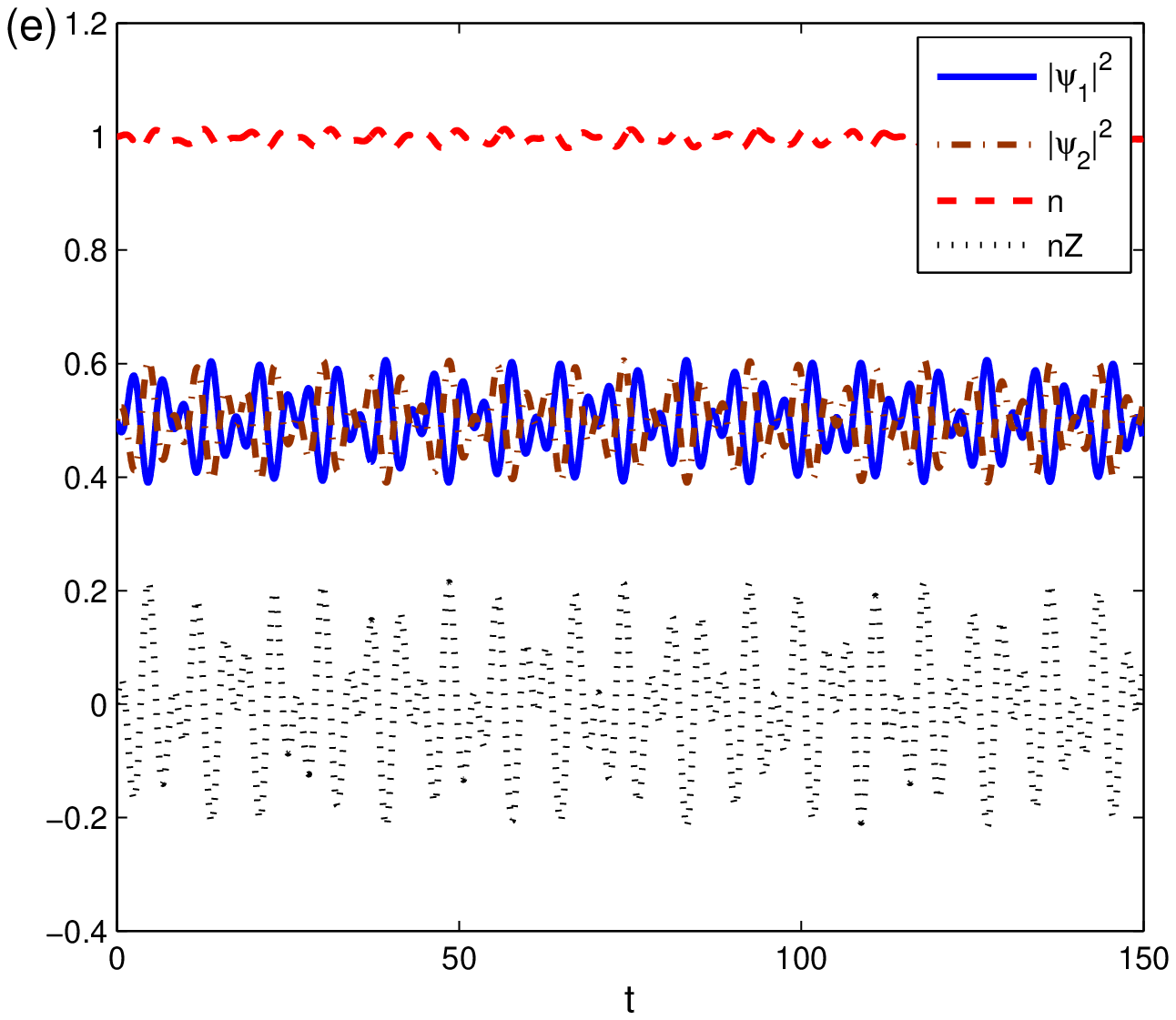}
\caption{Regular, periodic motion of a $\mathcal{PT}$ dimer with a small gain-loss strength, $\gamma/\nu=0.1$ and a strong nonlinearity, $\lambda/\nu=2$, driven by a field with an intermediate frequency $\omega/\nu=1$, but a small amplitude $A/\nu=0.2$.  The temporal evolution of $Z(t)$ (a) and $\theta(t)$ shows periodic behavior with multiple, closely space frequency components. The phase-space portrait (c) and the Poincare section (d) show that the dimer has a regular motion although it is close to boundary with the chaotic region. (e) The norm of the state $n(t)$, on-site weights, and their difference show oscillatory behavior indicative of the $\mathcal{PT}$ symmetric phase.}
\label{fig:four}
\end{figure}

\begin{figure*}
\centering
\includegraphics[width=\columnwidth]{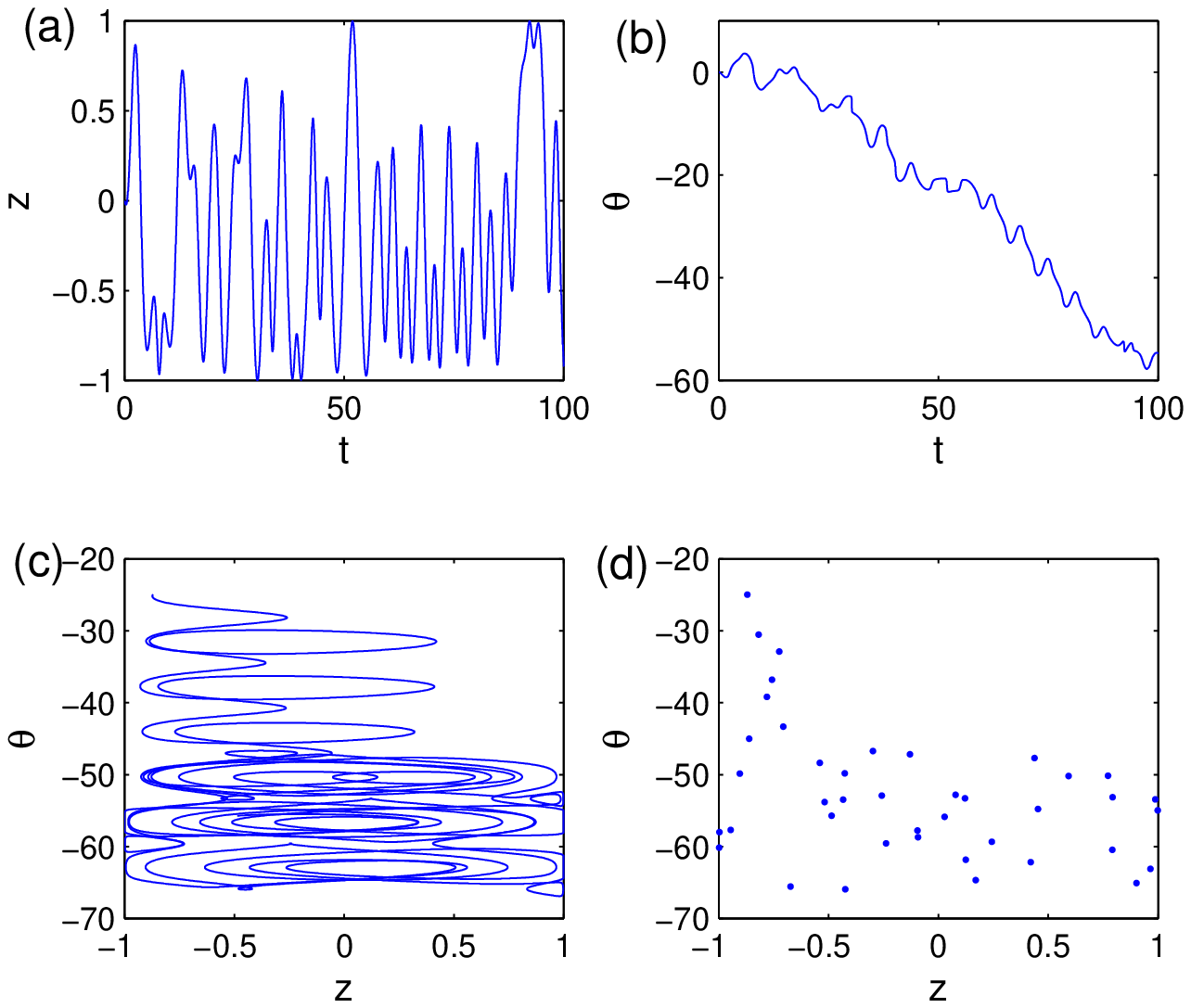}
\includegraphics[width=\columnwidth]{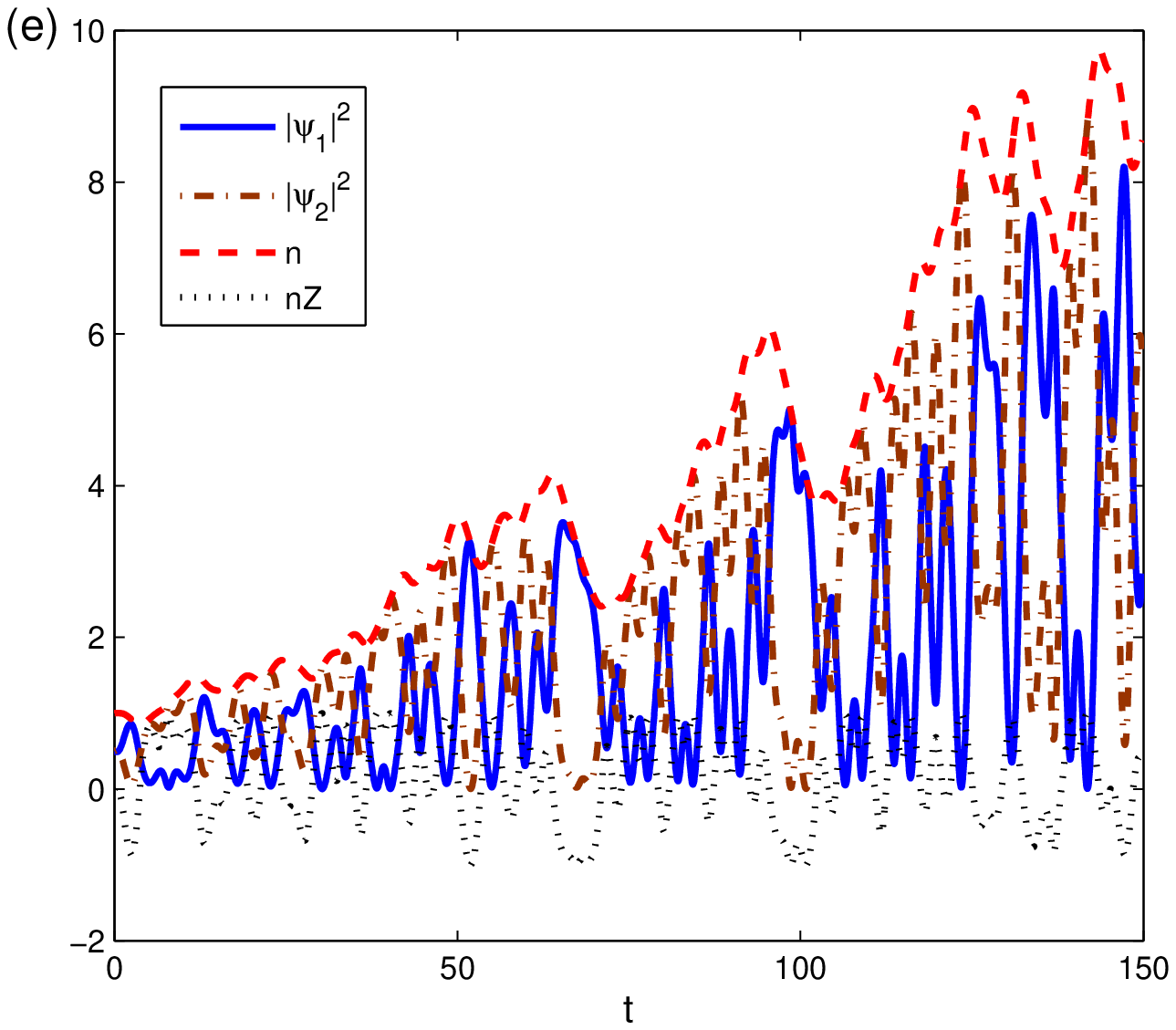}
\caption{Chaotic behavior of a $\mathcal{PT}$ dimer with a small gain-loss strength, $\gamma/\nu=0.1$ and a strong nonlinearity, $\lambda/\nu=2$, driven by a field with an intermediate frequency $\omega/\nu=1$ and a large amplitude $A/\nu=1.5$. $Z(t)$ (a) and $\theta(t)$ show aperiodic, oscillatory behavior. The phase-space portrait (c) and the Poincare section (d) show that the dimer is in the chaotic regime. (e) the norm of the state $n(t)$, on-site weights, and their difference show increasing-in-time behavior indicative of a $\mathcal{PT}$ symmetry broken phase.}
\label{fig:five}
\end{figure*}

To demonstrate regular and chaotic behavior at intermediate driving frequencies ($\omega/\nu=1$), we consider the $\mathcal{PT}$ symmetric dimer with a small and large drive amplitudes respectively, and obtain its time evolution with the same initial conditions as before. Thus, $\lambda/\nu=2,\gamma/\nu=0.1$ and $Z(0)=0=\theta(0)$. Figure 4 shows the results for a small $A/\nu=0.2$, when the dimer is in the regular region, but close to the boundary with the chaotic region (Fig.~\ref{fig:three}b). Panels (a)-(b) show that the dimer polarization and phase difference have an oscillatory behavior with closely spaced multiple frequencies $\sim\nu$. The phase-space portrait, Fig.~\ref{fig:four}c, and the Poincare section, Fig.~\ref{fig:four}d, show the regular motion of the dimer. Figure~\ref{fig:four}e shows that the norm of the state $n(t)$, the on-site weights $|\psi_{1,2}(t)|^2$, and their difference all oscillate, signaling a $\mathcal{PT}$ symmetric phase.

When the same dimer is driven with a stronger amplitude, $A/\nu=1.5$, it is in the chaotic regime. Figure~\ref{fig:five} shows the results for its temporal dynamics. The polarization $Z(t)$, Fig.~\ref{fig:five}a, and phase difference $\theta(t)$, Fig.~\ref{fig:five}b, do not show periodic behavior. The phase-space portrait, Fig.~\ref{fig:five}c, and the Poincare section, Fig.~\ref{fig:five}d, clearly show that the dimer is in deep, chaotic regime. Fig.~\ref{fig:five}e shows that the norm of the state $n(t)$, on-site weights $|\psi_{1,2}(t)|^2$, and their difference oscillate in an aperiodic manner and, on average, increase with time. This increase shows that the dimer is in the $\mathcal{PT}$ symmetry broken phase.

\section{Chaos vs. $\mathcal{PT}$ symmetry breaking}
\label{sec:chaos}

\begin{figure*}[tb]
\centering
\includegraphics[width=\columnwidth]{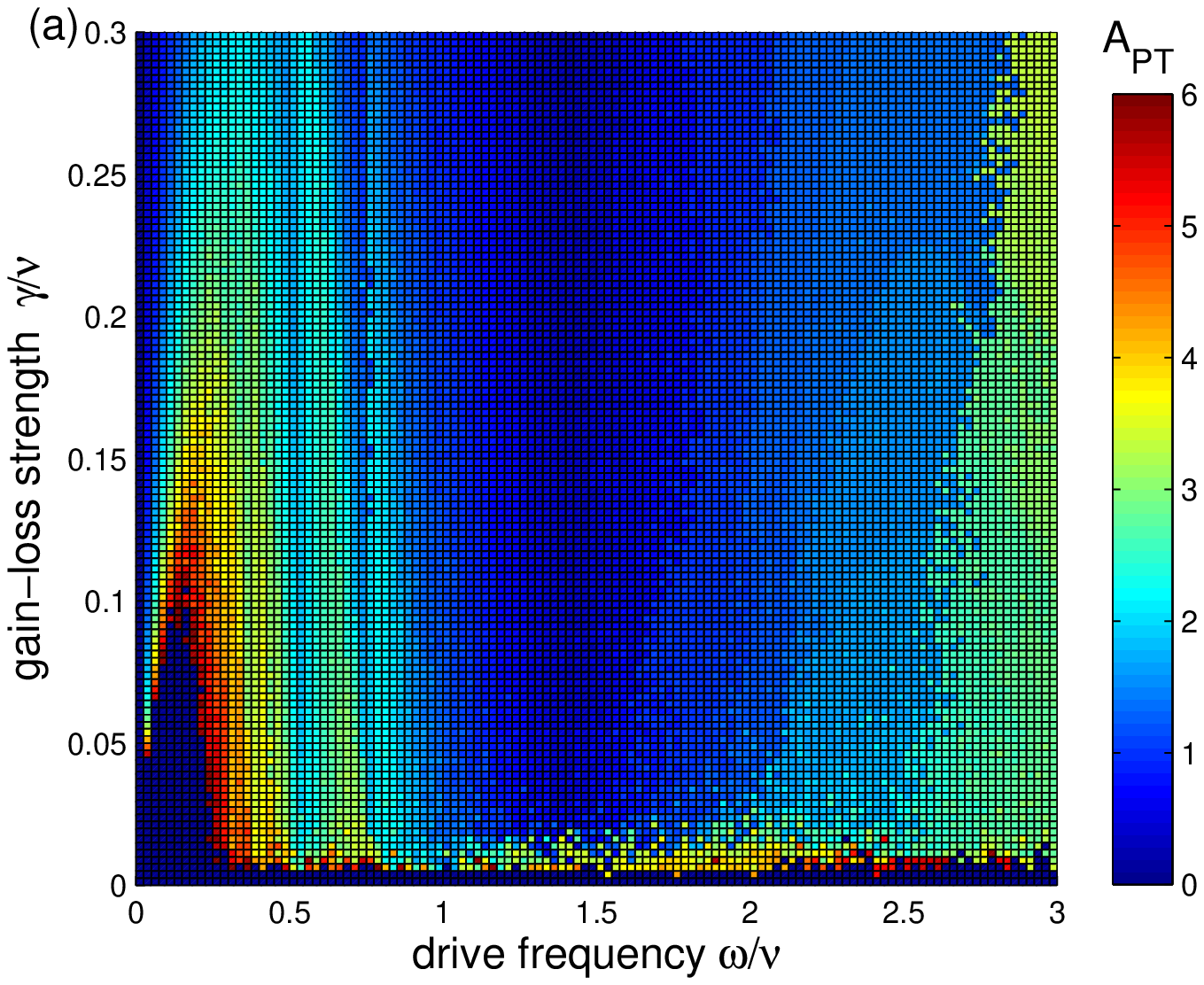}
\includegraphics[width=\columnwidth]{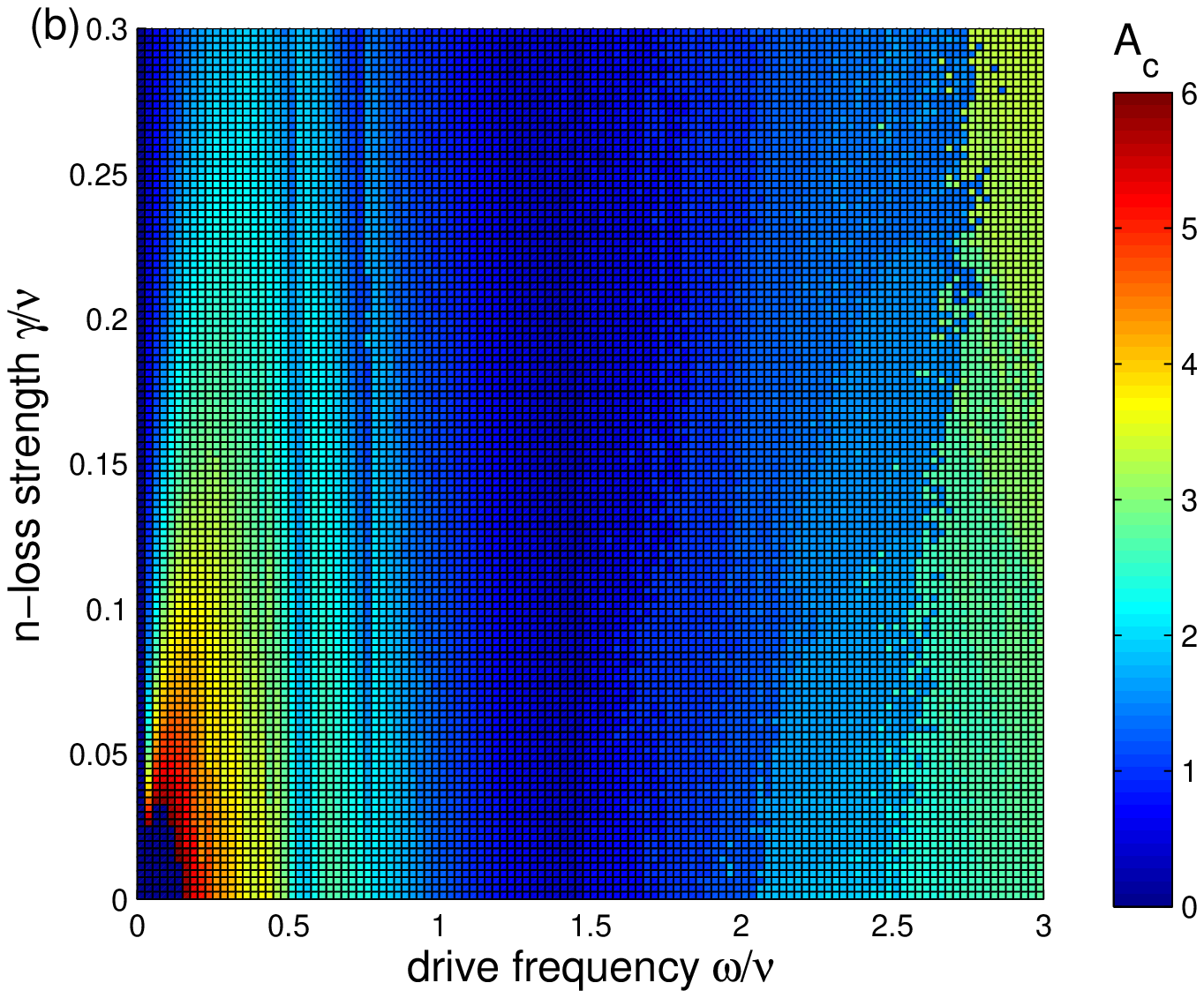}
\caption{The thresholds of $\mathcal{PT}$ symmetry breaking  and  chaos for $A/\nu$ in interval [0,6] which are represented by color. (a)The threshold of  $\mathcal{PT}$ symmetry breaking; (b) The threshold of  chaotic region. The zero value  means that it can't find $\mathcal{PT}$ symmetry breaking or chaos when $A/\nu$ in interval [0,6].}
\label{fig:six}
\end{figure*}

Numerical results from Sec.~\ref{sec:lowf},~\ref{sec:highf},~\ref{sec:inter} suggest a close correlation between regular dynamics and the $\mathcal{PT}$ symmetric phase, or chaotic dynamics and the $\mathcal{PT}$ symmetry broken phase. While chaos can be inferred from the phase-space portraits in the $Z-\theta$ plane, the definition of $\mathcal{PT}$ symmetry breaking in the absence of a static or Floquet Hamiltonian requires care. A key consequence of real spectrum of a non-Hermitian, $\mathcal{PT}$ symmetric system is bounded, oscillatory, non-unitary time evolution for the norm $n(t)$ of a state; a complex-conjugate spectrum, on the other hand, necessarily leads to unbounded-in-time dynamics for the norm. We use this criterion to define whether the dimer is in the $\mathcal{PT}$ symmetric phase or a $\mathcal{PT}$ symmetry broken phase.

For comparing the region of $\mathcal{PT}$ symmetry breaking and chaos, we set $\lambda/\nu=2$
and use the same initial state to obtain the $\mathcal{PT}$ symmetry breaking transition, threshold drive strength $A_{PT}/\nu$, Fig.~\ref{fig:six}a, and the chaos threshold drive strength $A_c/\nu$, Fig.~\ref{fig:six}b, while restricting to drive amplitudes $0\leq A/\nu\leq 6$; if the threshold is not found in this range, we assign it a zero value. The similarity between the two plots enforces the strong correlation between chaotic regime and $\mathcal{PT}$-symmetry broken regime in this system. The primary exception is the Hermitian case, i.e. $\gamma=0$, where the system is always in the $\mathcal{PT}$ symmetric phase, but can be driven to chaotic domain from a regular domain by increasing the drive amplitude. A detailed understanding of the differences between the chaotic/regular domains and $\mathcal{PT}$-symmetric/$\mathcal{PT}$-broken regions remains a topic of future work.

\section{Conclusion}
\label{sec:conc}

We have investigated the dynamics of a $\mathcal{PT}$ symmetric, nonlinear dimer under a periodic driving field. We have shown that while the motion remains regular and the dimer remains in the $\mathcal{PT}$ symmetric phase at low and high driving frequencies, both chaotic and regular dimer behavior emerges at intermediate driving frequencies. It is worth noting that the $\mathcal{PT}$-symmetry broken regions are strongly correlated with the chaotic region in the parameter space.

Controlling the dynamics through external fields is a subject of great interest, particularly so in open systems with balanced gain and loss. Our results provide new insights into the interplay between $\mathcal{PT}$-symmetry breaking transitions, that are caused by increasing the gain-loss strength, and the interaction-driven chaotic-to-regular transitions. They also raise the following key question: are these two, seemingly distinct transitions independent or not? How does one characterize the regions where the transitions are correlated or anti-correlated? A detailed investigation of these questions will deepen our understanding of interacting $\mathcal{PT}$ symmetric systems.

\begin{acknowledgments}
This work was supported by the National Natural Science Foundation of China under grant No. 11547019 and 1156011, China Scholarship Council (CSC,201609480009), and NSF grant No. DMR-1054020 (YJ). We thank Professor Chaohong Lee for fruitful discussions.
\end{acknowledgments}

\end{document}